\documentclass[10pt]{article}
\usepackage{mathrsfs}
\usepackage{multicol}
\usepackage{amsmath}
\usepackage{cite}
\usepackage{bm}
\usepackage{amsfonts}
\newcommand{\ct}[1]{{\textsuperscript{{\cite{#1}}}}}

\newcommand{\bee}{\begin{equation}}
\newcommand{\ee}{\end{equation}}
\newcommand{\beea}{\begin{eqnarray}}
\newcommand{\eea}{\end{eqnarray}}
\newcommand{\x}{{\bf x}}
\newcommand{\p}{{\bf p}}
\newcommand{\dpp}{{ \Delta{\bf p}^3\over (2\pi)^3}}
\newcommand{\vv}[1]{{\bf #1}}

 \newlength{\halfpagewidth}
 \divide\halfpagewidth by 2
                \newcommand{\leftsep}{%
                    \noindent\raisebox{4mm}[0ex][0ex]{%
                         \makebox[\halfpagewidth]{\hrulefill}\hbox{\vrule height 3pt}}%
                         \vspace*{-2mm}%
                }
                \newcommand{\rightsep}{%
                    \noindent\hspace*{\halfpagewidth}%
                         \rlap{\raisebox{-3pt}[0ex][0ex]{\hbox{\vrule height 3pt}}}%
                         \makebox[\halfpagewidth]{\hrulefill}%
  }

\begin{document}
\title{On wave functional in QED}
\author{{Daqing Liu}\\
        {\small Department of Applied Physics, Xi'an Jiaotong
        University, P.R. China}
       }
       \date{}
\maketitle
\begin{abstract}
In a discrete form of the second quantization, the gauge
independencies of all the physical states including vacuum in QED
are restudied through a new approach. We also discuss an interesting
phenomenon attributed to vacuum effect and come up with a procedure to produce general physical states. 
\end{abstract}

\vskip 0.1in

\begin{multicols}{2}
\section{Introduction}
Quantization of electromagnetic field theory is a textbook
task\ct{wheeler}. However, quantization using canonical formulation
faces two conflicting requirements: 1)the theory is gauge invariant;
2) there is lower bound to the free field energy. A obvious result
from the confliction is that we always introduce ghost states in
canonical quantization, such as gauge-dependent temporal photon and
longitude photon. As is known, one way to exclude the ghost states
is to take gauge fixing, for example, Coulomb gauge.  However, under
gauge fixings, the theory always loses gauge independence.
Furthermore, in temporal gauge, the gauge field is not fixed and
ghost states are still needed in canonical quantization.

Advanced studies show that there is possibly another way to exclude
ghost states, functional approach\ct{9710.3958}. For references we
refer to \cite{wheeler,ibb1,ibb2,9710.3958,9306161,ldq,lee} etc.
Refs. \cite{9710.3958,9306161,ldq} are on fermions while Refs.
\cite{wheeler,ibb1,ibb2} on the ground state of gauge fields.
However, in the functional approach, the same as in the canonical
quantization, the gauge independence of physical states is not very
obvious. Lee argues that all states should be gauge independent to
consist with a peculiar phenomenon, color confinement, in
QCD\ct{lee}. But such phenomenon does not occur in QED. Here we
propose a new approach which can ensure that all physical states in
QED should also be gauge independent.

This approach takes advantage of the fact that QED possesses an
expansion symmetry in gauge space, which is the generalization of
local gauge transformation. Since such symmetry is not held for
general wave functionals, it is natural to require that the energy
of physical wave functional is invariant under such transformation.
Such requirement leads to the gauge independence of wave functional
in QED.

To avoid divergences and ambiguities in continuous theory attributed
to infinite ultraviolet and infrared cutoff, we discretize the
position space by dividing the box with size $L^3$ into
$N(\rightarrow \infty)$ grids with spacing $\Delta x=\Delta y=\Delta
z={L\over N^{1/3}}$ to get a finite ultraviolet and infrared cutoff.
For instance, if we set $L\rightarrow \infty$, we shall obtain
divergent results in Eqs. (\ref{de1}), etc. Furthermore, if
space-time is indeed discrete and/or QED is invalid beyond some
energy scale, the discreteness will have physical meaning.

In section 2 we list results of quantization to free QED. We show
that under a reasonable assumption, all the physical states are
gauge independent in section 3. Section 4 studies state functional
including vacuum in detail. Section 5 is a simple discussion.

\section{The Quantization to free QED}
This section shows the main results of the quantization to free QED
briefly. For simplicity, we set $A_0\equiv 0$.

The commutation relations of gauge fields $A_i(\x)$ and adjoint
fields $\Pi_i(\x)$ read \bee
[A_i(\x),\Pi_j(\x^\prime)]={\mathrm{i}\over \tau}
\delta_{ij}\delta_{\x,\x^\prime}, \ee where $\tau=\Delta x^3$. In
other words, $\Pi_i(\x)=-\mathrm{i}\frac{\partial}{\tau\partial
A_i(\x)}$. Meanwhile, suppose Fourier transformations of gauge
fields and their conjugate fields are defined as \bee A_i({\bf
p})=\sum \tau A_i(\x)e^{-i{\bf p}\cdot \x}, \,\Pi_i({\bf p})=\sum
\tau \Pi_i(\x)e^{i{\bf p}\cdot\x} \ee respectively, then \bee
[A_i(\p),\Pi_j(\p^\prime)] = [A_i^*(\p),\Pi_j^*(\p^\prime)]=
\mathrm{i}L^3\delta_{ij}\delta_{\p\p^\prime}. \label{commu}\ee

We also introduce the magnetic fields $B_i(\x)=\epsilon_{ijk}
\partial_jA_k(\x)$, or,
$B_i(\p)=\mathrm{i}\epsilon_{ijk}\hat{p_j}A_k(\p)$, where $B_i({\bf p})=\sum
\tau B_i(\x)e^{-i{\bf p}\cdot \x}$ and $\hat{p_j}\equiv {1\over \Delta x}\sin
p_j\Delta x$ (thereinafter we always ignore the hat symbol without confusion).
Thus, for instance, with the notation $\dpp=1/L^3$, \bee \frac{\partial}{\tau
\partial A_i(\x)}=\mathrm{i}\epsilon_{ijk}L^3\sum_\p \dpp e^{-i\p\cdot\x}
p_j{\partial \over \partial B_k(\p)}. \ee

Since $B_i(\x)$ and $A_i(\x)$ are both real, state functionals are
invariant under transformation $A_i(\p)\rightarrow A_i^*(-\p)$ or
$B_i(\p)\rightarrow B_i^*(-\p)$.

We read Hamiltonian as\ct{lee}
\beea H&=&{1\over 2}\sum_x \tau[\Pi_i\Pi^*_i+B_iB^*_i] \label{hh1} \\
&=&{1\over 2}\sum_x \tau[-{\partial^2\over \Delta x^6
\partial A_i\partial A_i^*}+B_iB_i^*]. \label{hh2} \eea

Or, in Fourier space, \end{multicols}
                \leftsep
 \beea H&=&{1\over
2}\sum\dpp\{-L^6{\partial^2 \over \partial A_i(\p)\partial
A_i^*(\p)} +p^2A_i(\p)A_i^*(\p)- p_iA_i(\p)p_jA_j^*(\p)\}
\nonumber \\ &=&{1\over 2}\sum\dpp\{-L^6{p^2\partial^2\over
\partial B_k(\p)\partial B_k^*(\p)} +L^6{p_j p_k\partial^2\over
\partial B_k(\p)\partial B_j^*(\p)} +B_i(\p)B_i^*(\p)\}.
\label{eqhm}\eea \rightsep
                \begin{multicols}{2}

\section{Gauge independence of state functionals}
We show here the properties of state functionals under gauge
transformation.  Hamiltonian in equation (\ref{eqhm}) can be
divided into $H=\sum\limits_\p H_\p$, where \beea
   H_\p &=&{1\over 2L^3}\{-L^6{\partial^2 \over \partial
A_i(\p)\partial A_i^*(\p)}
+p^2A_i(\p)A_i^*(\p)- \nonumber \\ &&
p_iA_i(\p)p_jA_j^*(\p)\}. \eea

Therefore, equation $H\Theta=E\Theta$ possesses separable solutions
$\Theta=\prod\limits_\p \Theta_\p[{\bf A}(\p)]$, where $\Theta_\p$'s
satisfy \beea \label{eqhn} \{-L^6{\partial^2 \over
\partial A_i(\p)\partial A_i^*(\p)}
+p^2A_i(\p)A_i^*(\p)- \nonumber \\
p_iA_i(\p)p_jA_j^*(\p)\}\Theta_\p=2E_\p L^3\Theta_\p, \eea with the
total energy $E=\sum\limits_\p\dpp E_\p L^3=\sum E_\p$.

As for a definite $\p$, the theory is rotation invariance providing
$p\ll \Delta x^{-1}$
. One can, therefore, rotate vector $\p$ into $\p_0=(0,0,p)$. For
such $\p_0$ we get
 \end{multicols}
                \leftsep
\beea &&\{-L^6\sum\limits_i{\partial^2 \over
\partial A_i(\p_0)\partial A_i^*(\p_0)}
+p^2A_1(\p_0)A_1^*(\p_0)+A_2(\p_0)A_2^*(\p_0)\}\Theta_{\p_0}=2E_{\p_0}
L^3\Theta_{\p_0}. \label{eq9}\eea

Eq. (\ref{eq9}) also possesses separable solution
$\Theta_{\p_0}=X[A_1(\p_0),A_1(-\p_0)]Y[A_2(\p_0),A_2(-\p_0)]Z[A_3(\p_0),A_3(-\p_0)]$,
with $X,\,Y,\,Z$ satisfying \bee \left\{\begin{array}{c}
  \{-L^6{\partial^2 \over \partial A_1(\p_0)\partial
A_1^*(\p_0)}+p^2A_1(\p_0)A_1^*(\p_0)\}X=2E^XL^3X, \\
    \{-L^6{\partial^2 \over \partial A_2(\p_0)\partial
A_2^*(\p_0)}+p^2A_2(\p_0)A_2^*(\p_0)\}Y=2E^YL^3Y , \\
    -L^6{\partial^2 \over \partial A_3(\p_0)\partial
A_3^*(\p_0)}Z=2E^ZL^3Z , \\
\end{array} \right.\label{solt}\ee where $E_{\p_0}=E^X+E^Y+E^Z$.
               \rightsep \begin{multicols}{2}
Now we have divided $\Theta_{\p_0}$ into two parts. One of them, $X$
and $Y$, is perpendicular to gauge transformation, and the other,
$Z$, is parallel to gauge transformation.The perpendicular part
resembles harmonic oscillator while parallel part free particle.

As for physical state, $X(Y,Z)$ tends to zero when
$|A_i|\rightarrow \infty(i=1,2,3)$. For $X$ and $Y$, with analogy
to oscillator, there is no problem. But for $Z$, there is no
solution satisfying the condition. Up to a constant, the general
solution can be written as $Z=\exp\{a
A_3-2E^ZL^{-3}a^{-1}A_3^*\}$, where, for simplicity, $A_3$ and
$A_3^*$ stand for $A_3(\p_0)$ and $A_3^*(\p_0)$ respectively. But
this solution is not convergent when $|A_3|\rightarrow \infty$,
provided $aa^*\neq 2E^ZL^{-3}$.
States with $E^Z<0$ can also be ruled out by the divergence of
functional at $|A_3|\rightarrow \infty$. Meanwhile, the choice of
$|a|= \sqrt{2E^ZL^{-3}}$ ($E^Z\geq 0$) gives a finite but
non-vanishing $Z$ when $|A_3|\rightarrow\infty$. Since each
eigen-functional, including for $E^Z=0$, has such problem, we  take
a  modified constraint on $Z$: $Z$ is finite when $|A_3|\rightarrow
\infty$.

Thus we obtain $Z=e^{a A_3-a^*A_3^*}$ with $|a|= \sqrt{2E^ZL^{-3}}$.
Here $A_3$ or ${\bf p\cdot A}$ is free completely, correspondingly,
$\Pi_3$ or ${\bf p\cdot\Pi}$ is determined absolutely, which can
also be seen from the conservation of ${\bf p\cdot\Pi}$, $\,[{\bf
p\cdot\Pi},H]\equiv 0$. This is a special case of Heisenberg
Uncertainty Principle.

It easy to see from Eq. (\ref{hh1}) that the system has symmetry,
$A_i\rightarrow A_i-p_i f, \Pi_i\rightarrow \Pi_i$. The local gauge
symmetry corresponds to a translation in gauge space, since $f$ is
an arbitrary scalar function of $\p$. However, $f$ can also be a
scalar function with respect to gauge fields, for instance,
$f=p_iA_i\epsilon$, where $\epsilon$ is independent of
$A_i$.\footnote{ This corresponds to a transformation
$A_i(\x)\rightarrow A_i(\x)+\sum \Delta y^3 \frac{\partial A_j({\bf
y})}{\partial y_j} \frac{\partial h({\bf x-y})}{\partial x_i}$. The
transformation is not local in position space, but in Fourier space,
it is. } It is easy to check that under this transformation
$\vv{\Pi}$ and $\vv{B}$ remain unchanged. Therefore, besides local
gauge symmetry, QED also possesses an expansion symmetry in gauge
space.

However, unlike the local gauge symmetry, the expansion symmetry is
broken after Eq. (\ref{hh2}). To see it we perform a transformation
in gauge space, ${\bf A}\rightarrow {\bf A}+{\bf p}h$, where we
choose scalar function $h=\epsilon{\bf p\cdot A}/p^2$. At ${\bf
p}={\bf p}_0$ we get
 \bee\left\{
\begin{array}{c}
  A_3\rightarrow A_3+a\epsilon A_3, \\
  A_3^*\rightarrow A_3^*+a^*\epsilon^* A_3^*, \\
  A_1(A_2,A_1^*,A_2^*)(\p)\rightarrow A_1(A_2,A_1^*,A_2^*)(\p), \\
\end{array} \right. \ee
Then in equation (\ref{solt}) $Z\rightarrow
Z^\prime=e^{a(1+\epsilon)A_3-a^*(1+\epsilon^*)A_3^*}$. The new
functional has a changed energy, $E^{\prime Z}=|1+\epsilon|^2 E^Z$,
that is, the state has a energy of gauge dependence as long as
$E^Z\neq 0$. The statement can be considered in another way. As we
know, $Z$ is a functional with a (complex) period, which is in
proportion to $(E^Z)^{-1}$ up to a phase factor. Since the above
transformation changes the period of the functional, it can also
change $E^Z$.

We are faced with a puzzle: On one hand, $E^Z$ is a conservational
quantity, while on the other hand, it can be changed by an
unphysical expansion in gauge space. To treat this puzzle, reference
\cite{wbg} makes a gauge fixing, such as $A_3=0$, and no $\Pi_3$
existing correspondingly, for it is thought that neither $A_3$ nor
$\Pi_3$ has physical meaning, or, in other words, they are both
redundant variables at the case of $\p=\p_0$. This treatment takes
gauge dependent functionals and one should also modify the
commutation relation (\ref{commu}). Here we can treat it in another
way. We do not take the gauge fixing and therefore do not change the
commutation relation. On the contrary, we think that all the
physical states have a natural constraint: the energy of physical
state does not change under the gauge translations and gauge
expansions, since these transformations are both unphysical. This
requirement will lead to $E^Z=0$ and therefore $Z\equiv 1$.
Therefore, although there is no phenomenon similar to color
confinement, all the states should be gauge invariant in QED. For
general $\p$, the statement can be written as \bee
p_i\Pi_i\Theta=p_i\Pi^*_i\Theta=0. \label{gginv}\ee

The puzzle nominated as color confinement in QCD has been treated by
many researches, most of which are based on some combined forces.
For instance, in reference \cite{inj} the author introduces a
non-local Coulomb interactions between color charge. Here we show a
somewhat different viewpoint.

In non-Abelian case, especially $SU(3)$ theory or QCD, we face a
very tanglesome situation. In QED, interactions in Hamiltonian is
local in Fourier space (up to a $\pm \p$). However, they are
nonlocal in non-Abelian theory. This is because there occur cubic
and quartic interactions in QCD. An infinitesimal local gauge
transformation (in position space) connects different momentum and
color direction(A finite local gauge transformation even connects
states with different numbers of gluons). Consider a gluon with
single momentum (up to a $\pm \p$) and/or single color direction.
Suppose  it is an eigen-state of Hamiltonian in QCD, it can be
written as $B_i^a(\p)\Theta_0$ (or $A_i^a(\p)\Theta_0$), where
superscript and subscript are color index and direction index
respectively. The gluon will be connected with other gluons with
different momentum and/or directions, for instance,
$B^b_i(\p^\prime)\Theta_0$ (Generally, $|\p^\prime| \neq|\p|$) by
local gauge transformation, for the QCD vacuum $\Theta_0$ is gauge
invariant( This is a significant difference between Abelian theory
and non-Abelian theory). Therefore, since $B_i^a(\p)\Theta_0$ and
$B^b_i(\p^\prime)\Theta_0$ are connected by a local gauge
transformation, which does not change state energy, they have the
same energy. This is impossible unless single gluon with definite
momentum is infinite heavy. Or, a single gluon is eigenstate of
Hamiltonian if and only if it is infinite heavy.

Unlike in reference \cite{wbg}, our treatment keeps the commutation
relations in equation (\ref{commu}) unchanged and does not introduce
gauge condition. By a constraint on physical states, we find that,
attributed to the gauge expansion symmetry, not only vacuum, but
also all the physical states are gauge independent.

\section{Solution to state functional }
In this section we show the solution to general wave functional.
First we review the functional of vacuum, the eigen-functional
with the lowest energy.

At first, one possibly prefers writing the vacuum state as
functional with respect to $B_i$. But such treatment will meet a
singularity. To see it we write the vacuum state from equation
(\ref{eqhn}) as
 \bee \Theta_0= \exp\{-\sum\dpp
B_i^*(\p)D^0_{ik}(\p)B_k(\p)\}, \ee due to the translation
invariance. Introducing a positive matrix
$S^0(\p)=D^0(\p)+D^{0T}(-\p)$ we have
  \bee 1/p^2=S^0(1-\bar{P}/p^2)D^0, \ee where $(\bar{P})_{mn}=p_mp_n$.
There is no solution to this equation, for the determinant of
l.h.s. equal to $(1/p^2)^3$ while the determinant of r.h.s. equal
to zero, unless the determinant of matrix $S^0$ equals to
infinity.

To see it more clearly, we write $S^0={1\over
p}(1-\bar{P}/p^2)^{-1/2}$ naively. Suppose $\p=(0,0,p_3)$, or
$1-\bar{P}/p^2=diag(1,1,0)$,  we then obtain a singular $S^0_{33}$.
This reveals an obvious fact that there is no longitudinal magnetic
fields in free QED.

Therefore, a more convenient proposal is to write the vacuum state
as functional with respect to $A_i$, \bee \label{vf} \Theta_0=
\exp\{-\sum\limits_\p \dpp A_i^*(\p)D_{ik}(\p)A_k(\p)\}. \ee
Repeating the deductions, we obtain, \bee D={p\over
2}(1-\bar{P}/p^2). \ee

It is easy to check that $\Theta_0[{\bf A}(\p)]=\Theta_0[{\bf
A}(\p)+\p h]$, where $h$ is an arbitrary scalar function. As
expected, $\Theta_0$ is gauge independent.

Iterate $B_i(\p)=\mathrm{i}\epsilon_{ijk}p_jA_k(\p)$ into Eq.
(\ref{vf}), we write the vacuum functional as, \bee
\Theta_0=\exp\{-\sum\dpp {1\over 2p}B_i^*(\p)B_i(\p)\},
\label{18}\ee with a constraint $p_iB_i=p_iB^*_i=0$. This result is
in agreement with the references \cite{ibb1,ibb2}, except a
necessary constraint. Since canonical fields are ${\bf A}(\x)$, we
prefer (\ref{vf}) to (\ref{18}) as our final result.

For the density of the ground state energy, we have \bee
\mathcal{E}_0=E_0/L^3={1\over 2}\sum\dpp 2D_{ii}(\p)={d\over
2}\sum\dpp p, \ee where $d=3-1=2$ is just the degree of freedom.
Thus, due to the gauge invariance, the degree of freedom is not
three but two for each \({\bf p}\). In discrete form $\mathcal{E}_0$
is \bee \int_{-\pi\over\Delta x}^{\pi\over \Delta x}
\dpp\sqrt{\sin^2 p_x\Delta x+\sin^2 p_y\Delta x+\sin^2 p_z\Delta
x}\simeq{1.19\over \tau}. \ee The ultraviolet cutoff in the Fourier
space is just the inverse size of grids $\Delta x^{-1}$. In
continuous {\it l.h.s.} in the above equation should be $
\int_{-\pi\over\Delta x}^{\pi\over \Delta x} \dpp p$, which is about
630 times larger than the discrete one. It is significant that the
zero-point energy in the discrete form is much lower than that in
the continuous form.

The most possible measurements of canonical fields and their conjugate fields,
electric fields, are vanishing at each $\p$. However, other measurements are
still possible. This is uncertainty in quantum mechanism. In fact, any definite
configuration, such as ${\bf A}_i(\x)\equiv 0$, is never the eigenstate of
Hamiltonian, attributed to the uncertainty. Thus, if we put an electric dipole
in a box, its motion will be changed by nonzero electric field originated from
the uncertainty. Such effect is suppressed by the volume of box.

The uncertainty also leads to the condensates of the gauge fields.
Without loss of generality we set $\p_0=(0,0,p_3)$. We have now
$<A_3^*(\p_0)A_3(\p_0)>=\infty$ for the vacuum is gauge independent.
But, the condensates $<A_1^*(\p_0)A_1(\p_0)>=<A_2^*(\p_0)A_2(\p_0)>$
are finite: \beea &&<A_1^*(\p_0)A_1(\p_0)>= \frac{\int [d{\bf
A}(\p)]A_1^*(\p_0)A_1(\p_0) \Theta_0^2}{\int [d{\bf A}(\p)]
\Theta_0^2} \nonumber \\ &&= \frac{\int
dA_1(\p_0)A_1^*(\p_0)A_1(\p_0) \exp\{-{2p_0\over
L^3}A_1^*(\p_0)A_1(\p_0)\}}{\int dA_1(\p_0) \exp\{-{2p_0\over
L^3}A_1^*(\p_0)A_1(\p_0)\}} \nonumber \\
&&=\frac{L^3}{2p_0},\label{de1} \eea where $p_0=|p_3|$.

One can furthermore obtain the gauge independent condensates,
\beea &&<B_i^*(\p_0)B_i(\p_0)>= \nonumber \\&&
p^2_0<A_1(\p_0)^*A_1(\p_0)+A_2^*(\p_0)A_2(\p_0)>=\frac{2}{2}p_0L^3,
\nonumber \\ && <\Pi_i^*(\p_0)\Pi_i(\p_0)>=\nonumber
\\ &&-L^6<{\partial^2\over
\partial A_i^*(\p_0)\partial A_i(\p_0)}>=\frac{2}{2}p_0L^3.\label{cde}\eea
The gauge invariance implies $\Pi_3(\p_0)=0$ in the vacuum with a
completely free $A_3(\p_0)$. Notice here that not only gauge field
but also gauge potential, which is gauge dependent, has gauge
independent expectations. One can generalize Eq.
 (\ref{cde}) to general $\p$ in a straight way, which leads to an expected result, $E_0={1\over
2}\sum\dpp\{<B_i^*(\p)B_i(\p)>+<\Pi_i^*(\p)\Pi_i(\p)>\}$.

It is also interesting to study correlators of the gauge fields at
different positions. The results are \beea
<B_i^*(\x)B_i(0)>&=&{1\over L^3}\sum \dpp
<B_i(\p)B_i^*(\p)>e^{i\p\cdot \x} \nonumber \\
&=&\sum \dpp p\, e^{i\p\cdot\x}, \nonumber \\
<\Pi_i^*(\x)\Pi_i(0)>&=&\sum \dpp p\, e^{i\p\cdot\x}.\eea

The study shows that the vacuum proposes many properties similar to
that of the ground state in the harmonic oscillator, but there are
also some different properties, due to the gauge invariance of QED
vacuum. Because of the quantum effect, or, uncertainty, the vacuum
has a complex structure, for instance, the nonvanishing condensates
and correlate.

The following is a simple study on the general solutions. We
emphasize again that all the state must be gauge invariant. We take
solution to Eq. (\ref{eqhn}).

Setting $A_i^\perp\equiv A_j(\p)(\delta_{ij}-p_ip_j/p^2_0)$,
$\Theta_{\p_0}\equiv\Theta_{\p_0}[A_i^\perp,\,A_i^{\perp *}]=f\Theta_{0\p_0}$
and $\Theta_{0\p_0}=\exp\{-{p_0\over L^3}A_i^\perp A_i^{\perp *}\}$ with
function $f$ to be determined, we have, \beea \label{eqhn1} E_{\p_0}
f&=&-L^3(\delta_{ij}-p_ip_j/p^2_0) {\partial^2 f\over \partial A_i^{\perp
*}\partial A_j^\perp} +\nonumber \\ && p_0({\partial f\over
\partial A_i^\perp}A_i^{\perp}+{\partial f\over \partial
A_i^{\perp *}}A_i^{\perp *}) , \eea where we have
ignored the ground state energy.

One can use equation (\ref{eqhn1}) to study states of photons. For
instance, up to a constant, we obtain a quantum state
 \bee \Theta^k=(c_kA_k^\perp+c_k^*A_k^{\perp
*})\Theta_{0},\ee where $c_k=e^{i\theta_k}$ and
$\Theta_0=\prod\limits_{\hbox{pairs of $\p$}}\Theta_{0\p}$ is the
vacuum. Notice here $p_k\Theta^k\equiv 0$. It is not difficult to
verify that $\Theta^k$'s are two eigenstates with linear
polarization perpendicular to the momentum \(\p_0\).

One can use the skill of state superposition to construct the states
of photon corresponding to other direction of linear polarization or
corresponding to circular polarization. Furthermore, the study of
state of single photon can also be generalized to other states, for
instance, states of multi-photon.

\section{Discussions}
The gauge independencies of QED are studied through a new approach,
which is related to a generalization of the local gauge symmetry,
the expansibility in gauge space. The study shows clearly that all
the physical states are gauge independent.

Our study reveals clearly why there are just two degrees of freedom
in gauge field and therefore the introduction of ghost states is not
needed. Furthermore, we show that not only all physical states, but
also expectations of operators, some of which, for instance, gauge
potential, $A_i$, is not gauge independent, should be gauge
invariant.

All the states should be gauge invariant both in QED and in QCD.
Whereas, there is a crucial difference between QED and QCD, that is,
gauge particle, photon, exists in QED. We hope our approach be
helpful to understand color confinement in QCD.

\end{multicols}
\end{document}